# Hyperparameter Optimization and Force Error Correction of Neuroevolution Potential for Predicting Thermal Conductivity of Wurtzite GaN

Zhuo Chen[1,2], Yuejin Yuan[1], Wenyang Ding[3], Shouhang Li[4], Meng An[3*], Gang Zhang[2*],

[1]College of Mechanical and Electrical Engineering, Shaanxi University of Science and Technology, Xi'an, Shaanxi 710049, China

[2] Yangtze Delta Region, Academy of Beijing Institute of Technology, JiaXing, Zhejiang, China

[3]Department of Mechanical Engineering, The University of Tokyo, 7-3-1 Hongo, Bunkyo, Tokyo 113-8656, Japan.

[4]Centre de Nanosciences et de Nanotechnologies, CNRS, Université Paris-Saclay, 10 Boulevard Thomas Gobert, 91120 Palaiseau, France

**Corresponding author:**

anmeng@photon.t.u-tokyo.ac.jp (M. A.)

gangzhang2006@gmail.com (G. Z.)

**Abstract**

As a representative of wide-bandgap semiconductors, wurtzite gallium nitride (GaN) has been widely ulitized in high-power devices due to high breakdown voltage and low specific on-resistance. Accurate prediciton of wurtzite GaN thermal conductivity is a prerequiste for designing effective thermal mangament systems of electronic applications. Machine learning-driven molecular dynamics simulation offers a promsing approach to predicting the thermal condctivity of large-scale systems without requiring predefined parameters. However, these methods often underestimate the thermal conductivity of materials with inherently high thermal conductivity due to the large predicted force error compared with first-principle calculation, posing a critical challenge for their broader application. In this study, we successfully developed a neuroevolution potential for wurtzite GaN and accurately predicted its thermal conductivity, $259 \pm 6$ W/m-K at room temperatue, achieving excellent agreement with reported experimental measurements. The hyperparamters of neuroevolution potential (NEP) were optimzied based on systematic analysis of reproduced energy and force, structural feature, compuatational efficiency. Furthermore, a force prediction error correction method was implemented, effectively reducing the error caused by the additional force noise in the Langevin thermostat by extrapolating to the zero-force error limit. This study provides valuable insights and hold significant implication for advancing efficient thermal management technologies in wide-bandgap semiconductor devices.

## 1. Introduction

The wide-bandgap semiconductors, exemplified by GaN and SiC, have been widely ulitized in high-power devices due to high breakdown voltage and low specific on-resistance[1]. However, these high-power devices often operate under extreme conditions, such as high temperatures, high voltages, and high-frequency environments[2]. These conditions typically results in extremely high heat flux densities, which can advansely impact the devices' stability and operation lifespan[3,4]. Accurate prediction and characterization the thermal conductivity of wurtzite GaN are essential for understanding its thermal transport mechanisms and designing effectient thermal managment systems.

Over the past decades, extensive studies have focused on the prediction and characterization of the thermal conductivity of wurtzite GaN using both theoretical methods[5,6,7] and experimental measurements[8,9,10]. The measured thermal conductivity of wurtzite GaN typically has typically lacked a consistent conclusion, generally ranging from 180 to 220 W/m-K. This variability arises from differences in experimental samples, measurement methods, and the limited availiablity of single-crystaline wurtzite GaN samples. Recently, the development of high-pruity wurtzite GaN sample has enabled the measurement of thermal condcutivity up to 246 W/m-K through systematic experimental characterization[11]. In constrast, first-priple calculation-based Boltzmann transport equation tend to overestimate the thermal conductivity of wurtzite GaN using the local-density approximation (LDA) form of the echange-correlation funciton, primarily inconsistent lattice parameters with experimental samples and simplified anharmonicity calculations[5,12]. In constrast, our previous work[13] determined the thermal conductivity of wurtzite GaN to be 264 W/m-K, employing the PBE form of the exchange-correlation function while maintaining consistent lattice parameters (a = 3.184 Å and c = 5.187 Å). This results align well with experiment measurement[11]. Similarly, molecular dynamics (MD) simulations using emperical potential predicted a thermal conductivity of approximately 180 W/m-K[14,15], which is significantly lower than the reported experimental value[11]. Machine learning-driven molecular dynamics simulation offers a promsing alternative for predicting the thermal condctivity of large-scale systems without the need of predefined parameters. However, these methods often underestimate the thermal conductivity of materials with inherently high thermal conductivity[16,17], such as wurtzite GaN, posing a critical challenge for their broader application.

In this work, we first developed the neuroevolution potential (NEP)[18] for wurtzite GaN, a representative of machine learning potential (MLP) that combines high computational efficiency with excellent accuracy. We systematically inversigated the influence of NEP hyperparameters on the accuracy of thermal conductivity prediction and compuational efficiency. Furthemore, we implemented a correction method to eliminate force prediciton errors in the Langevin thermostat[19], achieving an accurate thermal conductivity of 259 ± 6 W/m-K at room temperatue, which aligns closely with experimental value[11].

## 2. Methods

### 2.1 Neuroevolution potential training

In NEP mode, the radial and angular components are influenced by the cutoff radii of $r_c^R$ and $r_c^A$, respectively. In our study, two cutoff radii of $r_c^R$ and $r_c^A$ were checked for optimize the accuracy of NEP model including the value of 4.5 Å, 5.0 Å and 5.5 Å. We performed tests to determine an optimal cutoff value, other hyperparameters used to train the NEP model are listed in Table S1. In total, we gathered 404 structures for the training set and 180 structures for the testing set. The training set includes an unitcell, a supercell, 200 MD structures at temperatures ranging from 100 to 1900 K, and 202 structures with tiny random atomic displacements and variations in lattice constants. We employed a general-purpose machine learning model, MACE-MP[20], in conjunction with the *Atomic Simulation Environment* (ASE) package[21] to perform MD simulations using the canonical (NVT) ensemble. The testing set was generated by performing MD simulations of heating from 100 to 1900 K using the NEP model, which was trained on the training set. Static calculations based on the first-principles method were performed to determine obtain the energy, atomic forces and virials of each structure.

### 2.2 DFT calculations for reference data generation

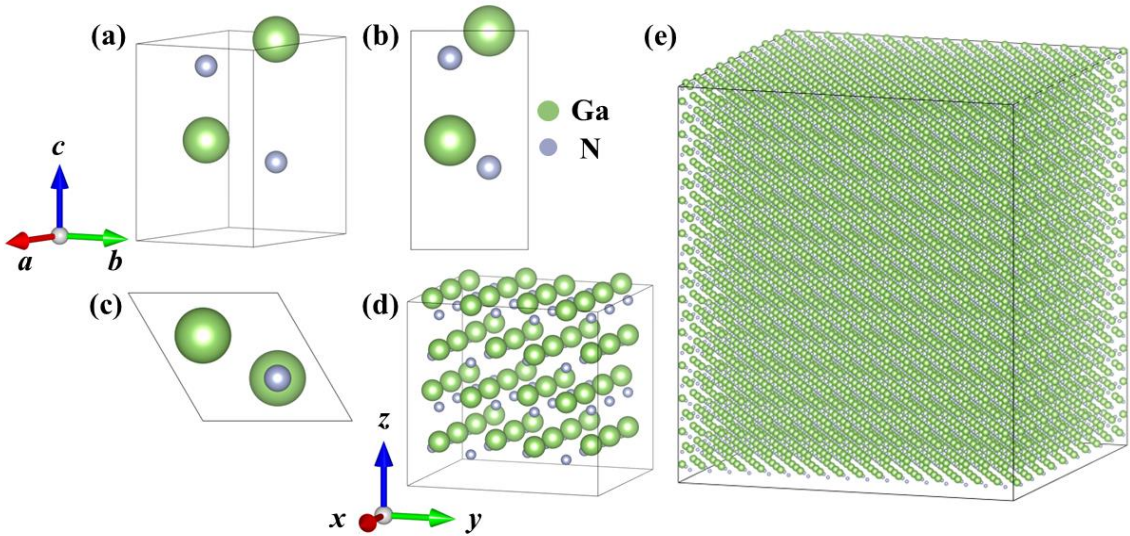

Figure 1. The snapshots of (a)-(c) the unitcell of wurtzite GaN, (d) the supercell for MACE-MP-based MD simulations, and (e) the supercell for thermal conductivity calculations.

First-principles calculations were performed using density function theory (DFT) as implemented in the *Vienna Ab Initio Simulation Package* (VASP)[22]. The unitcell of wurtzite GaN, as shown in **Figure 1(a-c)**, with a hexagonal system and the P6$_3$mc space group. The converged lattice constants of wurtzite GaN to be $a = b = 3.219$ Å and $c = 5.243$ Å, which are approximately 1% larger than the experimental values ($a = 3.19$ Å and $c = 5.189$ Å). This discrepancy is attributed to the PBE functional[23], which tends to overestimate lattice constants. In order to eliminate artificial factor, original lattice constants are not considered adjusted in this work. The hexagonal system was converted to a cubic lattice (**Figure 1(d)**) with dimensions 9.65 Å × 11.153 Å × 10.487 Å for MD simulations used to build datasets. Structural visualization was carried out using the VESTA package[24].

### 2.3 MD simulations

All MD simulations for thermal conductivity calculations were performed at 300 K using the GPUMD package[25] (version 3.9.4) with the NEP models (version 4)[26]. The simulation process involved three distinct runs, each with a time step of 1 fs. The first run, lasting 0.5 ns, employed a stochastic cell rescaling (SCR) thermostat[27] within the isothermal-isobaric (NPT) ensemble, with a target pressure of zero. The second run, conducted for 0.2 ns, used a Bussi-Donadio-Parrinello (BDP) thermostat[28] in the NVT ensemble. Finally, a 15 ns simulation was performed in the NVT ensemble, during which the running thermal conductivity was recorded. We only considered the thermal conductivity of wurtzite GaN along $z$-direction. The thermal conductivity was calculated using the homogeneous non-equilibrium molecular dynamics (HNEMD) method[29], which is unique to GPUMD. In this method, the heat current is proportional to the magnitude of the driving force parameter $F_e$, cautious selection of $F_e$ is essential to secure a large signal-to-noise ratio within the linear-response regime of the systems.

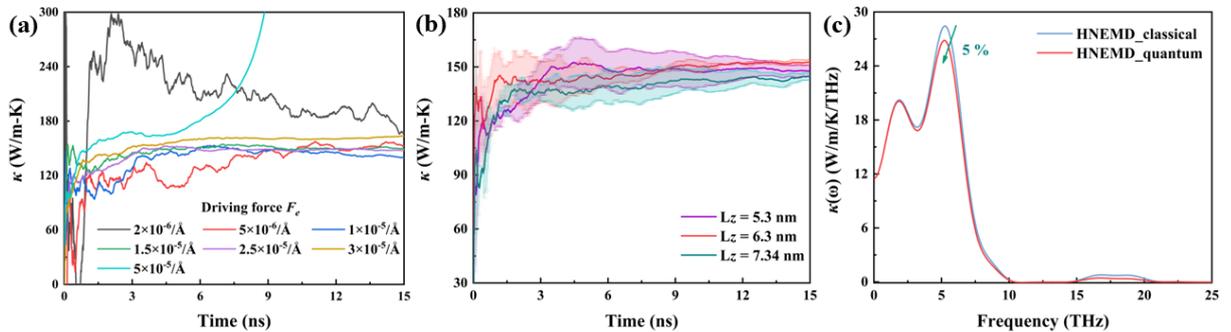

Figure 2. (a) Cumulative average of the thermal conductivity of wurtzite GaN at 300 K and zero pressure. The driving force parameter $F_e$ listed in the legend ranges from $2 \times 10^{-6}$ to $5 \times 10^{-3}$ Å$^{-1}$. (b) Cumulative average of the thermal conductivity of wurtzite GaN at 300 K and zero pressure. The length of $z$-direction increases from 5.3 nm to 7.34 nm, three independent runs

are performed. (c) The classical and quantum-corrected spectral thermal conductivity of wurtzite GaN at 300 K. The green arrow and number indicate the overvalued part of the classical MD.

To determine the parameter, we conducted extensive tests, the result is presented in **Figure 2(a)**. The system deviates from the linear-response regime when $F_e \geqslant 5 \times 10^{-5}$ /Å, while a small value of $F_e = 2 \times 10^{-6}$ /Å induces large fluctuations in thermal conductivity. Based on these observations, a value of $F_e = 2.5 \times 10^{-5}$ /Å was selected. Additionally, a sufficiently large simulation cell was used to eliminate finite-size effects. As shown in **Figure 2(b)**, we only considered the thermal conductivity of wurtzite GaN in the *z*-direction in this work, using a simulation cell with dimensions L*x* × L*y* × L*z* = 55.88 Å × 54.83 Å × 63.08 Å (depicted in **Figure 1(e)**), which containing 16320 atoms. In the classical MD simulations, vibrational modes follow the Maxwell-Boltzmann statistics, meaning each mode contributes to the heat capacity. However, in the quatunum statistical mechanics, only the vibrational modes with the frequencies below the cutoff $k_B T \leqslant \hbar\omega_0$, where $\omega_0$ is the vibrational frequency, contribute to the heat capacity, as described by Bose-Einstein statistics. To investigate the quantum effect, we employ a feasible quantum correction method based on spectral thermal conductivity is available within the HNEMD formalism[30]:

$$\kappa^q(\omega,T) = \kappa(\omega,T) \frac{x^2 e^x}{(e^x - 1)^2}, \qquad (1)$$

$$\kappa^q(T) = \int_0^\infty \frac{d\omega}{2\pi} \kappa^q(\omega,T). \qquad (2)$$

where $x = \hbar\omega/k_B T$, $\hbar$ is the reduced Plank constant and $k_B$ is the Boltzmann constant. $\kappa(\omega,T)$ is the classical spectral thermal conductivity and the total quantum corrected thermal conductivity $\kappa^q(T)$ is obtained as an integral of $\kappa^q(\omega,T)$ over the vibrational frequency. The quantum corrected spectral thermal conductivity at 300 K is presented in **Figure 2(c)**, we found that the thermal conductivity decreased by only 5% after quantum correction. This minimal change is attributed to the thermal conductivity of wurtzite GaN is primarily governed by low-frequency (<10 THz) phonons, which exhibit a weak quantum effect. As a result, quantum effect is not considered in the subsequent calculations.

## 3. Results and Discussion
### 3.1 Hyperparameter optimization of neuroevolution potential

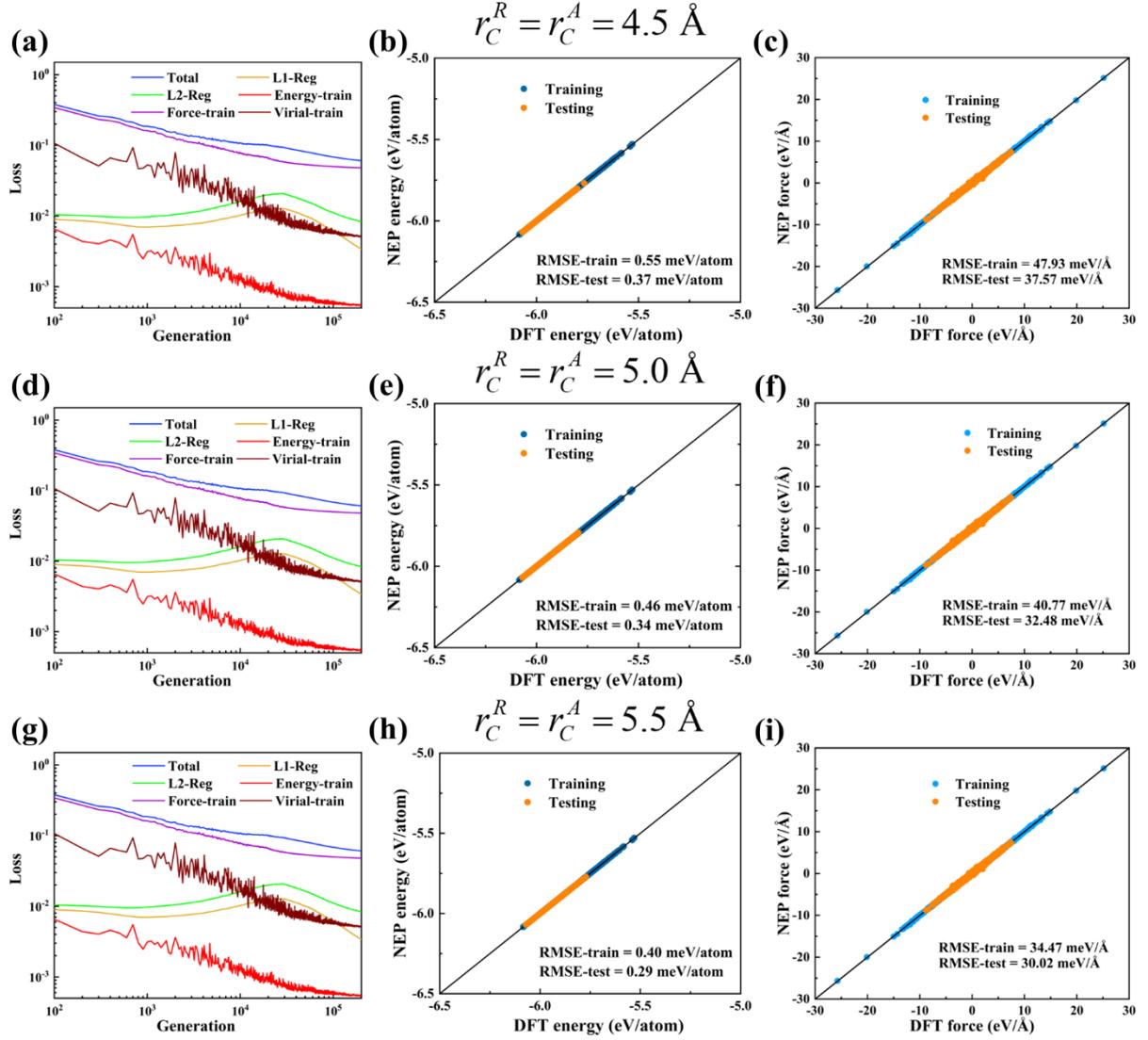

Figure 3. Evolution of the various terms in the loss function, including those for the L1 and L2 regularization, the energy RMSE (eV/atom), force RMSE (eV/Å) and virial RMSE (eV/atom) as a function of the training generation of (a) 4.5 Å, (d) 5.0 Å, and (g) 5.5 Å. Energy calculated from the NEP model as compared to the DFT reference data of (b) 4.5 Å, (e) 5.0 Å, and (h) 5.5 Å. Force calculated from the NEP model as compared to the DFT reference data of (c) 4.5 Å, (f) 5.0 Å, and (i) 5.5 Å. The overall converged training and testing RMSEs of energy and force are presented. Integrated cutoff values are indicated in above of each panel.

The NEP model is a MLP, based on the feedforward neural network and employed to describe the interatomic interaction. In a neural network potential, increasing the number of neurons generally improves the accuracy until convergence. Therefore, the hyperparameters such as the number of radial and angular descriptor components related to the number of neurons, often follow standard values. In MD simulations, when calculating short-range interactions, it is typically sufficient to account only for the contributions of particles within a cutoff distance. The introduction of truncation approximation greatly reduces the computational cost of interatomic interactions. However, the relationship between the accuracy and truncation of MLP is not monotonic, and there is usually an optimal cutoff that achieves the optimized training accuracy. Our calculations found that the NEP with the highest accuracy was associated with a significantly slower computational speed. Therefore, we question whether a greater cutoff radius is necessary to achieve higher accuracy. In previous study, the cutoff value was determined only after the accuracy test had been conducted[31]. In this work, we hope to do more comprehensive testing to determine a more reasonable cutoff value, which is very meaningful for the training process of standardizing NEP.

Figure 3 presents the training results of three NEP models with different cutoffs, including the radial ($r_c^R$) and angular ($r_c^A$) cutoffs. As depicted in **Figure 3(a)(d)(g)**, the loss curve plateaus as the number of training steps increases. The diagonal plots show the errors between the energy and force predicted by the NEP models and DFT calculations. The RMSEs for energy in the training set reveal a decreasing trend as the cutoff radius increases: 0.55 meV/atom for $r_c^R = r_c^A$ = 4.5 Å, 0.46 meV/atom for $r_c^R = r_c^A$ = 5.0 Å, and 0.40 meV/atom for $r_c^R = r_c^A$ = 5.5 Å. Similarly, the force RMSEs demonstrate a consistent improvement with increasing cutoff radii, decreasing from 47.93 meV/Å for $r_c^R = r_c^A$ = 4.5 Å, 40.77 meV/Å for $r_c^R = r_c^A$ = 4.5 Å, and 34.47 meV/Å for $r_c^R = r_c^A$ = 4.5 Å. These results indicate that enlarging the cutoff radius significantly enhances the accuracy of NEP model for capturing both energy and force of DFT calculation due to better representation of longer-range atomic interactions. Upon the optimization of energy and force in NEP, we then evaluated the predictive structure information of wurtzite GaN using the NEPs. All atomic structures were relaxed for 0.5 ns at 300 K using a SCR thermostat in the NPT ensemble, followed by 0.2 ns at 300 K using a BDP thermostat in the NVT ensemble. The radial distribution function (RDF) *g(r)* and the structural factor (SF) *S(q)* are shown in **Figure 4(a-b)**. SF highlights the long-range order characteristic of crystalline materials. Interestingly, the structural features remain consistent across different cutoff radii, indicating that a cutoff radii of 4.5 Å or larger is sufficient to accurately capture and construct the crystalline structure of wurtzite GaN.

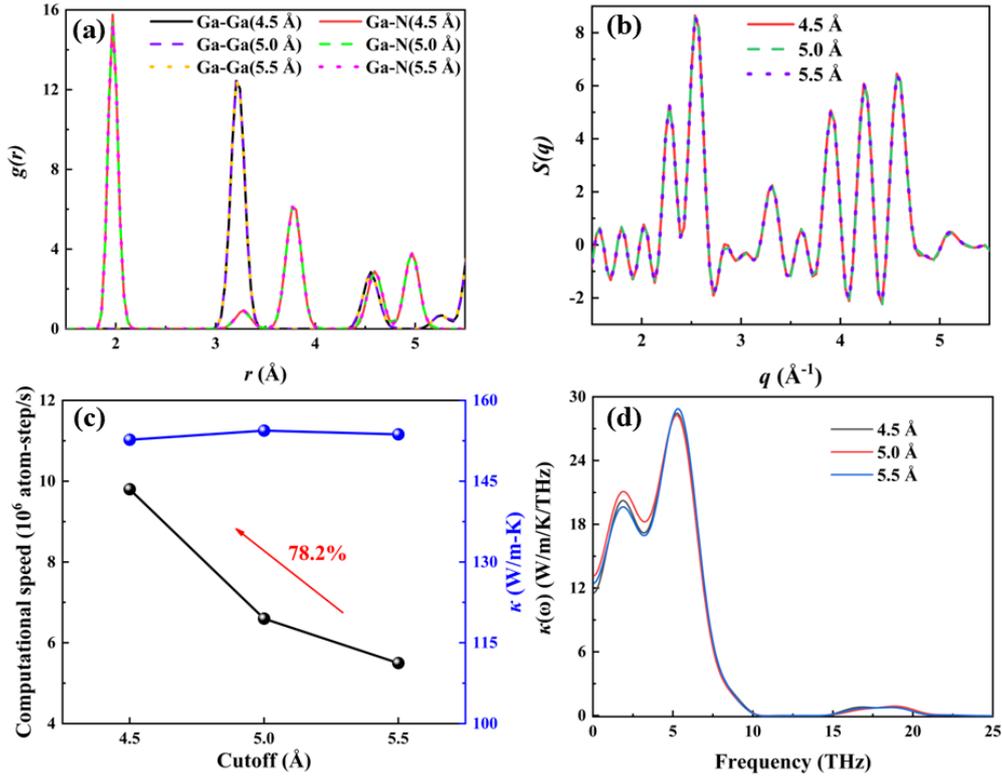

Figure 4. (a) Radial distribution function $g(r)$ of Ga-Ga and Ga-N predicted by NEPs with various cutoff radii. (b) The structure factors S(q) of wurtzite GaN. (c) The computational speed (black axis) of NEPs and the thermal conductivities (blue axis) of wurtzite GaN predicted by NEPs with various cutoff radii. (d) The spectral thermal conductivities predicted by NEPs with various cutoff radii.

In addition to evaluating the structure features using the NEP models, the computaional efficiency and accuracy of predicting the thermal conductivity of wurtzite GaN are equally critical. The thermal condcutivity of wurtzite GaN along the *z*-direction was calculated using NEPs with varied cutoff radii, as shown in **Figure 4(c)**. Notably, the thermal conductivity of wurtzite GaN remain consistent with differences of less than 2 W/m-K across three cutoff radii. In constrast, the computational speed with a cutoff radius of 4.5 Å is about $9.8 \times 10^6$ atom-step per second on a single Tesla A800 GPU, which is 78.2% faster compared to that with a cutoff radius of 5.5 Å. This significant improvement in computational efficiency highlights the advantage of adopting a smaller cutoff radius, which substantially reduces computational cost and time while maintaining accuracy in structural and thermal conductivity predictions. Furthermore, the spectral thermal conductivity obtained with varying cutoff radii shows consistent results across the entire range of phonon frequencies, as illustrated in **Figure 4(d)**. Based on the above systematic analysis of the reproduced energy and force, structure features, thermal conductivity predictions, computational efficiency and spectral thermal conductivity from NEPs with varying cutoff radii, a cutoff radius of 4.5 Å was selected to for subsequent thermal conductivity calculations, ensuring a balance between computational efficiency and accuracy.

**3.2 Correction of wurtzite GaN thermal conductivity**

Despite extensive paratermater optimization of the NEP models, the calculated thermal conductivity of wurtzite GaN at 300 K was around 150 W/m-K, which is significantly lower than the 246 W/m-K reported in experimental result[11] and the 260 W/m-K in DFT results[7,32]. Previous studies have highlighted the general tendency of MLPs to underestimate thermal conductivity, particularly in materials with inherently high thermal conductivity[16,17]. To address this limitation, researchers have developed correction method to reconcile the discrepancies between predicted and experimentally measrured thermal condcutivity[33,34]. In this study, we adopted the developed correction method to achieve a more accurate prediciton of thermal conductivity of wurtzite GaN.

The core of correction method lies in accouting for an additional force error, which follows a Gaussian distribution in the Langevin thermostat. This approach is analogous to the prediction error of forces observed in the NEP models, providing a systematic way to address discrepancies and enhance the accuracy of thermal conductivity prediciton based on MLPs. This is because the Langevin thermostat does not directly regulate the dynamics state of the entire system, as a global thermostat like BDP thermostat does. Instead, it introduces random forces of white noise to influence the dynamics of system, resulsting in an additional force error compared to a global thermostat. The magnitude of force error through the Langevin thermostat can be adjusted by the coupling time[33]:

$$\sigma_L^2 = \frac{2k_B T m_{ave}}{\tau \Delta t}. \tag{3}$$

where $m_{ave}$ is the average atom mass, $k_B$ is the Boltzmann constant, $T$ is the temperature in the system. $\tau$ is the coupling time and $\Delta t$ is the time step. The total force error $\sigma_{total}$ can be denoted as $\sigma_{total} = \sqrt{\sigma_{nep}^2 + \sigma_L^2}$, where $\sigma_{nep}$ is the RMSE of force prediction for the NEP model at the target temperature. The predicted thermal conductivity is directly proportional to $\sigma_{total}$, decreasing as $\sigma_{total}$ increases. Based on a first-order approximation and Matthiessen's rule, the corrected thermal conductivity $\kappa_0$ can be caculated using the follwing formula[33]:

$$\kappa_0 = \frac{1}{\frac{1}{\kappa} + \lambda \sigma_{total}}. \tag{4}$$

where $\kappa$ is the wurtzite GaN thermal conductivity by NEP-HNEMD with the langevin thermostat, $\lambda$ is a fitting parameter.

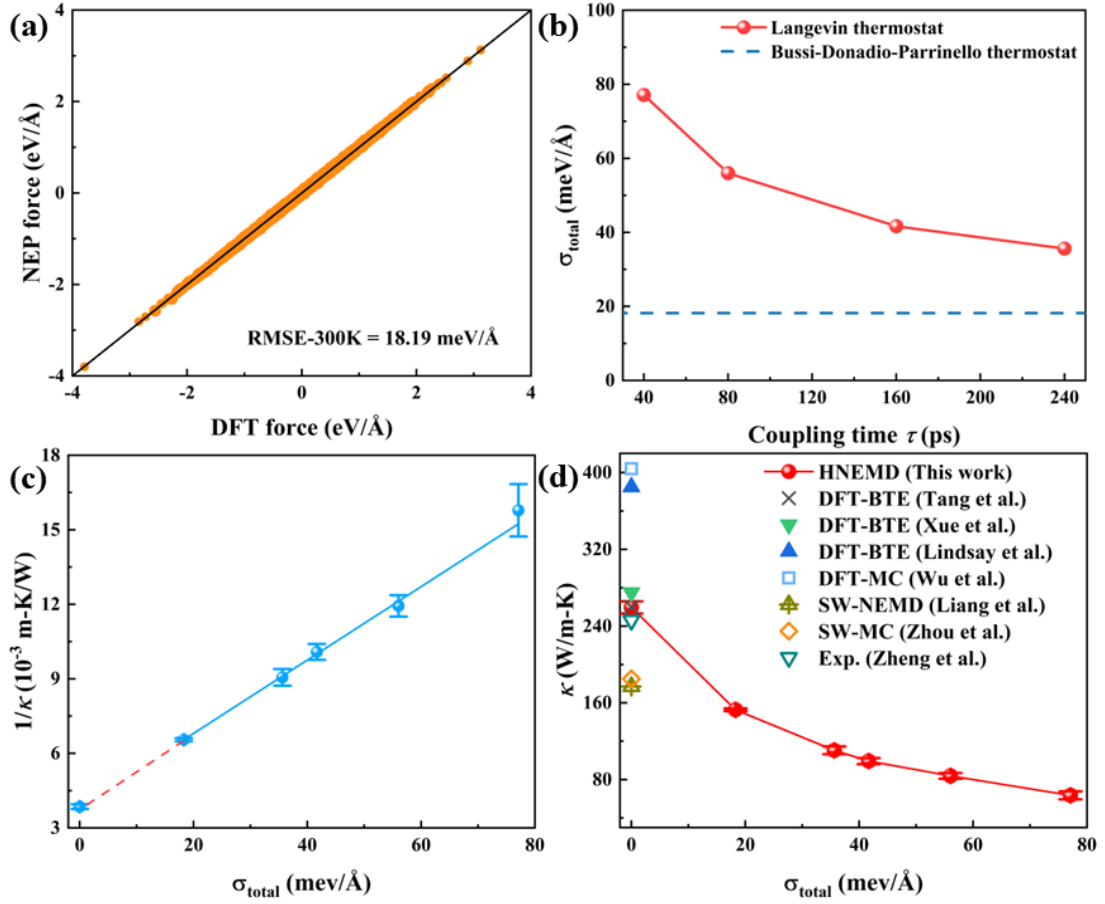

Figure 5. (a) Force prediciton of wurtzite GaN at 300 K from the NEP model compared with the DFT reference data. (b) Dependence of total force error $\sigma_{total}$ on the coupling time. Red dots represents $\sigma_{total}$ using the Langevin thermostat while blue dotted line repsresents $\sigma_{total}$ from the BDP thermostat, which remains unaffected by the coupling time. (c) Inverse thermal conductivity ($1/\kappa$) from NEP-MD simulations as a function of $\sigma_{total}$. The solid line represents linear fits, and the red dotted line is an extrapolation of the results. The intersection points at $\sigma_{total} = 0$ corresponds to the corrected thermal conductivity. (d) Calculated thermal conductivity as a function of $\sigma_{total}$. For comparsion, DFT values[5,7,12,32], SW results[14,15], and experimental data[11] of thermal conductivity of wurtzite GaN at 300 K are also shown.

We first evaluated the force prediciton error of NEP in the MD-simulated structures at 300 K by comparsing it to DFT reference data. These atomic configuration structures were generated through NEP-MD simulations using an SCR thermostat at 300 K, with 50 configurations uniformly sampled from 1 ns trajectory for DFT static calculations. **Figure 5(a)** shows the $\sigma_{nep}$ at 300 K, which can be considered as $\sigma_{total} = \sigma_{nep} = 18.19$ meV/Å when using the BDP thermostat. Notably, the force prediciton error of NEP model at 300K in **Figure 5(a)** is significantly lower than the overall force prediction error in **Figure 3(c).** This is because the training set also includes high-temperature and perturbed structures, which are typically more force-intensive and require the model to predicte the temperature-specific error accurately. Subsequently, we examined the force prediction error induced by the additional random force in Langevin thermostat by varying the coupling times to 240, 160, 80, and 40 ps.

**Figure 5(b)** illustrates the dependence of $\sigma_{total}$ on the coupling times of the Langevin thermostat, while $\kappa_0$ is determined in **Figure 5(c)** by fitting the data using Eq. (4). Using the fitted linear parameter, the thermal conductivity of wurtzite GaN was obtained by extrapolating to $\sigma_{total} = 0$. The uncorrected and corrected thermal conductivity of wurtzite GaN, obtained from NEP-MD simulations along with experimental data, DFT results, and SW-MD simulations results were presented in **Figure 5(d)**. The calculated thermal conductivities of pure wurtzite GaN from Lindsay et al.[5] and Wu et al.[12] are significantly higher than experimental value[11] due to the difference of lattice constant in wurtzite GaN. The uncorrected thermal conductivities are consistently lower than the other results due to the presence of force prediction errors in the NEP-MD calculations. Interestingly, the corrected thermal conductivity of wurtzite GaN from NEP-MD simulations is 259 ± 6 W/m-K, which aligns well with the experimental value of 246 W/m-K. The slight overestimation in our results may be attributed to the weak quantum effect at room temperature. Overall, we have achieved an accurate prediction of the pure wurtzite GaN thermal conductivity based on NEP-MD simulations. This work broadens the applicability of froce error correction method to a wider range of materials Furthermore, our finding confirms that wurtzite GaN does not exhibit significant isotopic effects, in contrast to earlier theoretical works.

## 4. Conclusions

In conclusion, we successfully developed a neuroevolution potential for wurtzite GaN and accurately predicted its thermal conductivity, achieving excellent agreement with reported experimental measurements. This work effectively addresses the common underestimate issue associated with high thermal conductivity prediction of materials using machine learning potential. The influence of the hydroparameters of two cutoff radii in NEP models on reproduced energy and force, structure features, computational efficiency and spectral thermal conductivity was systematically analyzed and the optimized cutoff radius of 4.5 Å achieves a balance between computational efficiency and accuracy. Furthermore, the force prediction error correction method of NEP models effectively reduces the error arising from the additional force noise in the Langevin thermostat by extrapolating to the zero-force error limit. This study provides valuable insights and hold significant implication for advcing efficient thermal management technologies in wide-bandgap semiconductor devices.

## Acknowlegement

This work was supported by the National Natural Science Foundation of China (No. 52376063 and No 52306116), the Hebei Key Laboratory of Low Carbon and High Efficiency Power Generation Technology Prevention Fund (Grant No. 2022-K03) and the China Postdoctoral Science Foundation (No. 2023MD744223).